# Whole-Earth Decompression Dynamics

J. Marvin Herndon

Transdyne Corporation
San Diego, California 92131 USA

June 30, 2005

## Abstract

The principles of *Whole-Earth Decompression Dynamics* are disclosed leading to a new way to interpret whole-Earth dynamics. *Whole-Earth Decompression Dynamics* incorporates elements of and unifies the two seemingly divergent dominant theories of continental displacement, plate tectonics theory and Earth expansion theory. Whole-Earth decompression is the consequence of Earth formation from within a Jupiter-like proto-planet with subsequent loss of gases and ices and concomitant rebounding. The initial whole-Earth decompression is expected to result in a global system of major primary decompression cracks appearing in the rigid crust which persist as the basalt feeders for the global, mid-oceanic ridge system. As the Earth subsequently decompresses, the area of the Earth's surface increases by the formation of secondary decompression cracks, often located near the continental margins, presently identified as oceanic trenches. These secondary decompression cracks are subsequently in-filled with basalt, extruded from the mid-oceanic ridges, which traverses the ocean floor by gravitational creep, ultimately plunging into secondary decompression cracks, emulating subduction. Much of the evidence presented in support of plate tectonics supports *Whole-Earth Decompression Dynamics*, but without necessitating mantle convection/circulation or basalt re-cycling. Moreover, the timescale for Earth decompression is not constrained to the last 200 million years, the maximum age of the current ocean floor.

## Introduction

For more than a century, scientists have recognized that opposing margins of continents fit together in certain ways and display geological and palaeobiological evidence of having in the past been joined. In the nineteenth century, Suess[1] envisioned a vast palaeocontinent that had broken up and partially subsided into the ocean floor. Suess's concept was subsequently recognized as untenable because continental rock is less dense than ocean floor basalt. Early in the twentieth century, Wegener[2] also proposed that the continents at one time had been united, but subsequently had separated and drifted through the ocean floor to their present positions. Wegener's theory of continental drift, generally not accepted for half a century, was revived in the 1960s, mainly as the result of investigations of oceanic topography[3-6] and palaeomagnetism[7-9], and modified to become plate tectonics theory.

Widely accepted at present despite certain unfounded fundamental assumptions, plate tectonics theory is predicated upon the idea that ocean floor, continuously produced at mid-oceanic ridges, moves like a conveyer belt, ultimately being subducted and re-circulated by assumed convection currents in the mantle[10-12]. Indeed, compelling evidence, *e.g.*, seafloor magnetic striations, exists to support the idea of seafloor being continuously produced at mid-oceanic ridges, moving away from the ridges and being subducted, for example, by entering oceanic trenches. To date, however, there is no direct, unambiguous evidence that mantle convection and/or mantle circulation actually takes place; in fact, there is some evidence to the contrary[13]. Moreover, there is no evidence that oceanic basalt can be repeatedly recycled through the mantle without being substantially and irreversibly changed. Yet, mantle convection/circulation and basalt recycling are fundamental necessities for the validity of plate tectonics. Furthermore, plate tectonics theory does not in and of itself provide an energy source for geodynamic activity.

In 1933, Hilgenberg[14-15] first published his observation that, on a sphere with a radius of about one half of the Earth's present radius, the continents more-or-less fit together like pieces of a jigsaw puzzle and, notably, form a uniform, continuous shell. Ideas that, at some time in the past, the Earth's radius was significantly less than its present value and that the separation of the continents occurs as a consequence of volume expansion of the Earth have been discussed for more than half a century[16-20]. As an alternative to plate tectonics theory, classical Earth expansion theory, as espoused by Carey[21] and others, has met with resistance because of the lack of knowledge of an energy source of sufficient magnitude and because it is predicated upon the idea that Earth expansion occurs mainly along mid-oceanic ridges and occurred during the last 200 million years, as the oldest ocean floor is no older than that[22]. Moreover, some argue, geological folding observed in various places, including in many mountain ranges, appears indicative of lateral stress rather than vertical uplifting.

Neither plate tectonics theory nor classical Earth expansion theory, I submit, is an adequate description of the dynamics of the Earth as a whole. Here, I propose a strikingly new theory that reconciles certain elements of those two seemingly divergent theories into one unified theory, designated *Whole-Earth Decompression Dynamics*. I propose that whole-Earth dynamics, as viewed from the surface, is characterized primarily by the following two distinct, but related, processes: (i) the formation of decompression cracks (often near continental margins), and (ii) the in-filling of those cracks with basalt (produced by volume decompression in the mantle), which is extruded mainly at mid-oceanic ridges, solidifies and traverses the ocean floor by gravitational creep to regions of lower gravitational potential energy, ultimately plunging downward into distant decompression cracks, thus emulating subduction[23]. As shown below, much of the evidence presented in support of plate tectonics supports *Whole-Earth Decompression Dynamics*, but without necessitating mantle convection/circulation. Moreover, the timescale for Earth decompression is not constrained to the last 200 million years, the maximum age of the current ocean floor.

### Protoplanetary Origin of *Whole-Earth Decompression Dynamics*

Planets generally consist of concentric shells of matter, but there has been no adequate geophysical explanation to account for the Earth's non-contiguous crustal continental rock layer, except by assuming that the Earth in the distant past was smaller and subsequently expanded. The principal impediment to the idea of Earth expansion has been the lack of knowledge of a mechanism that could provide the necessary energy[24-25] without departing from the known physical laws of nature[26]. In 1982, Scheidegger[27] stated concisely the prevailing view, "Thus, if expansion on the postulated scale occurred at all, a completely unknown energy source must be found." Recently, I disclosed just such an energy source that follows from fundamental considerations[23].

The so-called "standard model" of Solar System formation, which assumes the condensation of dust from a tenuous nebular and subsequent agglomeration into successively larger grains, pebbles, planetesimals and then planets is wrong and leads to the contradiction of terrestrial planets having insufficiently massive cores. Instead, I have shown the consistency of Eucken's 1944 concept[28] of planets raining out in the central regions of hot, gaseous protoplanets[29]. The Earth, together with its complement of lost primordial gases, comprises a protoplanetary mass remarkably similar to the mass of Jupiter. I have shown that formation of Earth, from within a Jupiter-like protoplanet, will account for the compression of the rocky Earth to about 64 percent of its current radius, yielding a closed, contiguous continental shell with whole-Earth decompression commencing upon the subsequent removal of its protoplanetary gaseous shell[23]. Gravitational energy of compression, stored during the Jupiter-like protoplanetary stage, may be seen as the primary energy source for subsequent Earth decompression during its approach toward a new hydrostatic equilibrium. To a much lesser extent, nuclear fission energy and radioactive decay energy[30-32] may augment the stored energy of protoplanetary compression, heating the interior of the Earth, and compensating to some extent for the cooling that results from decompression.

### Principles of *Whole-Earth Decompression Dynamics*

After being stripped of its great overburden of volatile protoplanetary constituents,

presumably by the high temperatures and/or by the violent activity, such as T Tauri-phase solar wind[33-36], associated with the thermonuclear ignition of the Sun, the Earth would inevitably begin to decompress, to rebound toward a new hydrostatic equilibrium. One may envision the initial rebounding to result in a global system of major primary decompression cracks appearing in the rigid crust as the internal shells expand, decompressing to lower densities. The locations of initial *primary* decompression cracks are identified presently as the basalt feeders for the global, mid-oceanic ridge system.

As the Earth subsequently decompresses and swells from within, the deep interior shells may be expected to adjust to changes in radius and curvature by plastic deformation. As the Earth decompresses, the area of the Earth's surface increases by the formation of *secondary* decompression cracks often located near the continental margins and presently identified as oceanic trenches. These secondary decompression cracks are subsequently in-filled with basalt, extruded from the mid-oceanic ridges, which traverses the ocean floor by gravitational creep.

Whole-Earth decompression should have commenced promptly upon removal of protoplanetary hydrogen and other volatile constituents. However, the timescale for the Earth's full rebounding from protoplanetary compression may be long, even extending into the present; witness, for example, the relatively minor rebounding of northern land masses, following post-Pleistocene deglaciation, being measured in thousands of years[37]. A much, much longer time may be expected for rebounding from compression due to protoplanetary-scale loading by approximately 300 Earth masses of volatile constituents. The timescale for decompression may be related to the pre-degasification protoplanetary thermal state, to the dynamics of degasification, especially the cooling that might have been involved, to mantle properties, to the cooling that results from decompression, and to the time required to replace heat lost by decompression cooling.

The Earth appears to be approaching the terminus of its decompression. If the Earth is presently decompressing, length of day measurements should show progressive lengthening. Such measurements, made with increasing precision over the last several decades, show virtually no current lengthening[38], implying no current secondary decompression crack formation. The formation of secondary decompression cracks might be episodic, though, like the release of stress by major earthquakes, or secondary crack formation may have ended forever. But major decompression cracks are still conspicuously evident, for example, circum-pacific trenches, such as the Mariana Trench. The complementary *Whole-Earth Decompression Dynamics* process of basalt extrusion and crack in-filling, however, continues at present.

Secondary crack formation and the in-filling of those cracks are complementary elements of the same *Whole-Earth Decompression Dynamics* process. Even in the absence of current secondary decompression crack formation, an estimate may be obtained of recent-period Earth decompression by considering the amount of in-filling basalt presently being produced. One may make the rough assumption that, over an arbitrary period of time, the rate of secondary crack volume formation at the Earth's surface is equal to the volume rate of extruded basalt. The volume rate of extruded basalt is thus related to the rate of decompression. The high-end estimate of 15 km$^3$/yr for current basalt production leads to an estimated annual percent increase in radius of only about 5 X 10$^{-10}$, a value not inconsistent with length of day measurements. Much higher basalt extrusion rates undoubtedly have occurred in the past, as the present estimated annual percent increase in radius, if constant over the lifetime of the Earth, would have only resulted in a 2 percent increase in radius.

## Geological Features

Plate tectonics, despite certain unfounded underlying assumptions, has enjoyed wide acceptance because it appears to describe many of the geological features that result from *Whole-Earth Decompression Dynamics*. There are subtle and profound differences, though, as shown by the following brief remarks on global and regional geological features.

The principal surface manifestation of the *Whole-Earth Decompression Dynamics* is the in-filling of secondary decompression cracks, located mainly near continents, with basalt extruded from mid-oceanic ridges. The following observations of oceanic features and the consequences of down-plunging slabs, usually arrayed as supporting plate tectonics theory, are, I submit, consequences of *Whole-Earth Decompression Dynamics*: ocean floor magnetic striations, transform faults, island arc formation, guyot formation, contributions of seamounts to coastal geology, and the generation and distribution of earthquakes. These have the same basis and understanding in *Whole-Earth Decompression Dynamics* as in plate tectonics. But there are global, fundamental differences between *Whole-Earth Decompression Dynamics* and plate tectonics, especially as pertains to the growth of ocean floor, to the origin of oceanic trenches, to the fate of down-plunging slabs, and to the displacement of continents.

In *Whole-Earth Decompression Dynamics* mid-oceanic ridges are thought to be the sites of the original, global system of primary decompression cracks which serve as persistent extrusion-basalt feeder-channels. There is no evidence that the mid-oceanic ridge system represents the edges of mantle-convection cells as implied by plate tectonics. Ancillary basalt extrusion also occurs from hot-spots, non-mid-oceanic-ridge feeders, such as those responsible for the formation of Hawaii and Iceland. Hot-spot basalt may ultimately fill secondary decompression cracks or it may add to continental mass directly or as incorporated terranes.

In *Whole-Earth Decompression Dynamics*, oceanic trenches, such as the Mariana Trench and others that rim the pacific basin, are thought to be surface manifestations of decompression cracks, more or less continuously being formed as the Earth decompresses and, notably, continuously being in-filled. Indeed, mantle seismic tomography results can be interpreted as imaging in-filled decompression cracks. There is no evidence that oceanic trenches represent the edges of mantle-convection cells as implied by plate tectonics and no reason to believe that the more-or-less lateral motion of moving seafloor can in and of itself produce trenches. In *Whole-Earth Decompression Dynamics*, oceanic troughs are thought to be partially in-filled decompression cracks.

Since Suess[1], understanding the process of mountain building has been hampered by conflicting evidence in a complex geological framework[39] and by limitations imposed through incorrect theories. For example, plate tectonics, while allowing for the development of lateral stress, is capable of admitting only asymmetric uplift by plate underthrust. Similarly, classical Earth expansion allows for symmetric uplift, but not lateral stress. In *Whole-Earth Decompression Dynamics*, on the other hand, both processes are possible. Lateral stress occurs for the same reasons as in plate tectonics and, additionally, occurs as a consequence of the formation of secondary decompression cracks. Symmetric uplift may also occur during decompression and asymmetric uplift by plate underthrust.

## Grand Overview

There has long been a duality of thought on the nature of the circumstances that gave rise to the planets of our Solar System. In 1944, on the basis of thermodynamic considerations, Eucken[28] suggested core-formation in the Earth, as a consequence of successive condensation from solar matter, on the basis of relative volatility, from the central region of a hot, gaseous protoplanet, with molten iron metal first raining out at the center. For a time hot, gaseous protoplanets were discussed[40-42], but emphasis changed abruptly with the publication by Cameron[43] of his diffuse solar nebula models at pressures of about $10^{-5}$ bar.

During the late 1960s and early 1970s, the so-called "equilibrium condensation" model was contrived and widely promulgated[44]. That model was predicated upon the assumption that the mineral assemblage characteristic of ordinary chondrite meteorites formed as condensate from a gas of solar composition at pressures of about $10^{-5}$ bar. During those years, scientists almost universally believed that the Earth is like an ordinary chondrite meteorite. Consequently, the idea that dust agglomerated into grains, pebbles, rocks and then into planetesimals seemed, superficially at least, to explain planet formation and planet composition. The problem, as I have shown, is that "equilibrium condensation model" is wrong[45] and the "standard model" of solar system formation would result in the terrestrial planets having insufficiently massive cores, a profound contradiction to observations[29]. Massive cores are instead indicative of terrestrial-planet-composition-similarity to enstatite chondrite meteorites, whose highly-reduced state of oxidation may be thermodynamically stable in solar matter only at elevated temperatures and pressures[29,46,47].

Rather than the presently popular idea of Earth formation solely from the accumulation of small bodies subsequent to the dissipation of hydrogen and other volatiles into space, I have suggested[29], as has Eucken[28] and others, that, prior to the ignition of thermonuclear reactions in the Sun, the proto-planetary Earth formed as a giant gaseous Jupiter-like proto-planet. This concept is more consistent with observations of close-in gas giants in other planetary systems[48]. Such a giant proto-planetary-Earth would be expected to have a differentiated core of enstatite-chondrite-like alloy-plus-rock overlain by nearly 300 Earth-masses consisting of hydrogen, noble gases, and other volatile constituents. Under such a great overburden of gases and ices, the relatively non-volatile chondrite-like alloy-plus-rock constituents, which now comprise most of the Earth, would be compressed by gravity to about 64% of the present day Earth radius[23].

Decompression of the Earth may be seen as a direct consequence of the subsequent removal of hydrogen and other volatile constituents, presumably during the thermonuclear ignition of the Sun. After being stripped of such a great overburden, the Earth would rebound, tending toward a new hydrostatic equilibrium. Gravitational energy of compression, stored during the Jupiter-like proto-planetary stage, may be seen as the primary energy source for driving geotectonic activity, augmented to a much lesser extent by nuclear fission and radioactive decay energy[49].

The initial whole-Earth decompression is expected to result in a global system of major primary decompression cracks appearing in the rigid crust which persist as the basalt feeders for the global, mid-oceanic ridge system. As the Earth subsequently decompresses, the area of the Earth's surface increases by the formation of secondary decompression cracks, often located near the continental margins, presently identified as oceanic trenches. These secondary decompression cracks are subsequently in-filled with basalt, extruded from the mid-oceanic ridges, which traverses the ocean floor by gravitational creep, ultimately plunging into secondary decompression cracks, emulating subduction. This is *Whole-Earth Decompression Dynamics*.